\documentclass[amsmath,amssymb,twocolumn,english,aps,manuscript,prl,reprint,10pt,A4,longbibliography]{revtex4}
\usepackage[T1]{fontenc}
\usepackage[latin9]{inputenc}
\usepackage{amsmath}
\usepackage{amssymb}
\usepackage{graphicx}

\begin{document}

\title{Tuning spin orbit interaction in high quality gate-defined InAs one-dimensional channels}

\author{J.~Shabani$^{1}$}

\author{Younghyun~Kim$^{2}$}

\author{A.~P.~McFadden$^{3}$}

\author{R.~M.~Lutchyn$^{4}$}

\author{C.~Nayak$^{2,4}$}

\author{C.~J.~Palmstr{\o}m$^{1,3,5}$}

\affiliation{$^{1}$California NanoSystems Institute, University of California, Santa Barbara, CA 93106, USA
\\
$^{2}$Department of Physics, University of California, Santa Barbara, CA 93106, USA
\\
$^{3}$Department of Electrical Engineering, University of California, Santa Barbara, CA 93106, USA
\\
$^{4}$Microsoft Research, Station Q, University of California, Santa Barbara, CA 93106, USA
\\
$^{5}$Materials Department, University of California, Santa Barbara, CA 93106, USA
}

\date{\today}

\maketitle
\textbf{Spin-orbit coupling in solids describes an interaction between an electron's spin, an internal quantum-mechanical degree of freedom, with its linear momentum, an external property. Spin-orbit interaction, due to its relativistic nature, is typically small in solids, and is often taken into account perturbatively. It has been recently realized, however, that materials with strong spin-orbit coupling can lead to novel states of matter such as topological insulators and superconductors~\cite{ Wilczek'09, Stern'10, Franz'10, Nayak'10}. This exciting development might lead to a number of useful applications ranging from spintronics to quantum computing~\cite{Kane_review, Qi_review, Alicea12, Beenakker_review}. In particular, theory~\cite{Sau10, Alicea10,  Lutchyn10, Oreg10} predicts that narrow band gap semiconductors with strong spin-obit coupling are a suitable platform for the realization of Majorana zero-energy modes, predicted to obey exotic non-Abelian braiding statistics~\cite{NayakRevModPhys}. The pursuit for realizing Majorana modes in condensed matter systems and investigating their exotic properties has been a subject of intensive experimental research recently~\cite{Mourik2012, Rokhinson2012, Das2012, Deng2012, Fink2012, Churchill2013}. Here, we demonstrate the first realization of gate-defined wires where one-dimensional confinement is created using electrostatic potentials, on large area InAs two dimensional electron systems (2DESs). The electronic properties of  the parent 2DES are fully characterized in the region that wires are formed. The strength of the spin-orbit interaction has been measured and tuned while the high mobility of the 2DES is maintained in the wire. We show that this scheme could provide new prospective solutions for scalable and complex wire networks.}

One-dimensional (1D) semiconducting quantum wires with strong spin-orbit interaction represent a unique platform for realizing Majorana zero-modes~\cite{Lutchyn10,Oreg10}. Epitaxial growth of two-dimensional heterostructures containing InAs layers is hypothesized to be a suitable system since InAs has strong spin-orbit coupling and large g-factor. In addition, Fermi level pinning in the conduction band at the surface of InAs could be used to control the transparency of InAs-superconductor interfaces, which is a necessary condition for Majorana devices.
Recent experiments on MOCVD grown self-assembled nanowires have shown great progress in revealing signatures of zero-energy Majorana bound states~\cite{Mourik2012, Rokhinson2012, Das2012, Deng2012, Fink2012, Churchill2013}. However, the next step in controlled assembly of nanowire networks~\cite{AliceaBraiding} using a bottom-up approach for scaling and building complex architectures poses a challenge for self-assembled nanowires.  Molecular beam epitaxy (MBE) growth of large area two-dimensional systems (2DESs) combined with semiconductor processing techniques in a top-down approach could provide a new avenue toward complex architectures. Fabrication of gate-defined devices on these systems is highly desirable as it offers the possibility of tuning confinement potential, carrier density, and spin-orbit coupling.

The spin-orbit interaction in 1D has been studied in self-assembled InAs nanowires \cite{HansenPRB05} and etched quantum wires \cite{SchaperPRB06}. Large spin-orbit values have been measured through weak antilocalization (WAL). However, the latter
effect is suppressed in narrower wires because the associated suppression of the diffusion constant causes WAL to evolve into
weak localization (WL) \cite{HansenPRB05, SchaperPRB06, Kunihashi10, RoulleauPRB10, KallaherPRB13}.

The strength of spin-orbit coupling is crucially important for the stability of an emergent topological phase in semiconductor-superconductor heterostructures since the energy gap is proportional to the spin-orbit coupling strength~\cite{Sau_review}. Furthermore, the detrimental effect of disorder on the topological phase must be minimized. Indeed, it has been shown~\cite{Motrunich01_Disorder_in_topological_1D_SC, Gruzberg05_Localization_in_disordered_SC_wires_with_broken_SU2_symmetry, Brouwer11_Topological_SC_in_disorder_wires, LobosPRL12} that impurity scattering, which is particularly important in one-dimensional structures, drives a quantum phase transition from the Majorana-carrying superconducting phase to a trivial phase with no Majorana zero-modes in the nanowire. Hence it is important to have both large spin-orbit coupling and high mobility in the semiconductor nanowires. It has been estimated theoretically that in order to observe Majorana zero modes, wire mobilities should be larger than $10^{5}$ cm$^{2}$/Vs \cite{Sau12, LobosPRL12}.  Thus, growth of a high-mobility 2DES on different crystal orientations and quantum well asymmetries would allow for both engineering of spin-orbit coupling strength and preserving high electron mobilities in gate-defined wires.

Gate-defined one-dimensional wires have been realized in GaAs 2D quantum wells and have led to major developments, such as sensitive spin qubits \cite{Petta05} and exotic interferometry devices \cite{Heiblum,Willett09a}. Unlike in GaAs, reliable gating has proven difficult in InAs due to gate leakage and hysteretic behavior. In addition, charge traps and Fermi level pinning could screen the applied electric field and significantly reduce the gate efficiencies. These difficulties
are surmounted in the present work, in which we grow high-mobility InAs heterostructures using MBE and demonstrate the fabrication of clean wires using surface gates.

\section*{Experimental approach}

High mobility InAs quantum wells have typically been grown on closely lattice matched antimonide barriers (GaSb, AlSb) \cite{Kroemer04}. These structures typically have high carrier densities which are difficult to deplete using surface gates. An alternate approach to growing InAs quantum wells on lattice matched antimonide barriers is the pseudomorphic growth of a thin strained layer of InAs on arsenide barriers (InGaAs, InAlAs). Relatively low electron densities could be achieved using this approach due to the different line-up of the impurity levels in InAlAs (compared to AlSb) barriers \cite{Capotondi04}. In this work we focus our effort on this latter approach. The 2DES is realized in an unintentially doped structure, grown on a semi-insulating InP (100) substrate using a modified V80H molecular beam epitaxy system. We start with growing a superlattice of In$_{0.53}$Ga$_{0.47}$As and In$_{0.52}$Al$_{0.48}$As at $T_{sub}$ = 480 $^{o}$C on InP substrate. Then the substrate temperature is lowered to $T_{sub}$ = 360 $^o$C and In$_{x}$Al$_{1-x}$As buffer is step-graded from $x=$ 0.52 to 0.8 and then stepped down to $x$=0.75. The electrons are confined to a 4 nm strained InAs layer inside an In$_{0.75}$Ga$_{0.25}$As quantum well \cite{Richter00}. The quantum well and top barriers are grown at a substrate temperature of $T_{sub}$ = 420 $^o$C \cite{Wallart05}. The As flux has been adjusted to achieve a (2 $\times$ 3) surface reconstruction during the step-graded buffer layer growth and a (2 $\times$ 4) for the In$_{0.75}$Al$_{0.25}$As and the channel. For samples with optimized low-temperature buffer layers, we have achieved mobilities over $3 \times 10^{5}$ cm$^{2}$/Vs. Figure 1 shows the schematic layout of the structure and the corresponding band-edge diagram calculated using a self-consistent Schrodinger-Poisson solver. In order to avoid gate-leakage and increase gate-tunability, the structure is not capped with In$_{0.75}$Ga$_{0.25}$As where signatures of parallel conduction (and artifically higher mobility values) have been observed. Instead, the sample is capped with SiO$_{x}$.

The measured mobility as a function of density at 2 K is shown in Fig.~1(c). This dependence is well-fitted by $\mu \sim n^{\alpha}$,  with $\alpha = 0.8$ at high density range ($n > 2 \times 10^{11}$ cm$^{-2}$). A value of $\alpha \sim 0.8$ signifies that the mobility is limited by scattering from nearby background charged impurities~\cite{DasSarmaHwangPRB13}. This is consistent with the low value of the critical density $ n_{c} \sim 3 \times 10^{10}$ cm$^{-2}$ we find for the metal-insulator transition. The density dependence of the conductivity ($\sigma \sim (n-n_{c})^{\delta}$) exhibits an exponent of $\delta = $1.79 ($\delta \sim 1+\alpha$), and indicates that the localization is caused simply by carrier trapping at defects ~\cite{DasSarmaHwangPRB13}.  Figure 1(d) shows normalized magnetoresistance ($R(B)/R_{B=0}$) measurements performed as the density is varied in a gated-Hall bar at 2.5 K. The gate stability of the device allows us to probe the evolution of integer quantum Hall states as a function of density and map the Landau fan diagram.

\section*{Gate-defined wires}

One-dimensional channels could be fabricated using surface split-gates where each gate creates lateral depletion and a 1D channel forms. Quantized plateaus observed in point contact geometries are hallmarks of ballistic transport in a 1D system \cite{vanWees88,Wharam88}. The quantum point contact (QPC) structure is also an essential building block for many other types of structures in mesoscopic physics, such as rings and quantum dots. Since the first observation of conductance quantization in QPCs on GaAs based structures \cite{vanWees88,Wharam88}, the quantization has been reported in electron systems in several other materials such as SiGe \cite{SiGe2}, GaN \cite{GaN}, and AlAs \cite{AlAs}. For a long time, there has also been interest in ballistic 1D transport in systems such as InAs with strong spin-orbit interaction and the resulting spin-splitting of the energy bands at finite wavevectors \cite{Winkler03}. The realization of high-quality and stable QPCs in these systems, however, has been challenging. Earlier experiments on InAs showed quantization of conductance, however, a background conductance had to be substracted \cite{Koester96}.

We use a split-gate design to fabricate point contacts and wires. Applying a negative DC voltage bias, $V_{sg}$, to the side gates depletes the electrons under the gates. Once all electrons are depleted, a quasi-one-dimensional channel forms with a sudden drop of conductance (an example is shown in Fig.~2). Full depletion is achieved in our devices at $V_{sg}$ = -3.3 V and less than 10 pA of leakage current is measured at $V_{sg}$ = - 3.5 V. In the supplement, we show devices with point contact geometry where transport is ballistic and conductance exhibits a quantized plateau at (2e$^{2}$/h) and a feature near 0.7(2$e^{2}/h)$ typically observed only in high quality GaAs quantum point contacts \cite{Cronenwett02, dePicciotto05}. We focus here only on the longer wires where transport is diffusive. In our device,  we could characterize the 2DES properties in the same region where the wire is formed, which is a unique feature of the gate-defined wires. Figure 2b shows magneto-transport measurements at $V_{sg}=0$ V. Near zero magnetic field (Fig.~2a), we observe a weak-antilocalization peak in the magneto-conductance due to the presence of spin-orbit coupling in our 2DES. The WAL is most useful as a tool in extracting spin-orbit parameters in moderately-disordered systems. Empirically, the WAL signal weakens and disappears as the mobility of the 2DES increases. This is mainly due to the vanishing probability of return in the interference paths for high-mobility samples. Here, we present data on a moderately lower mobility sample with $\mu \sim 6.5 \times 10^{ 4}$ cm$^{2}$/Vs at $V_{sg}$ = 0V.




The data is analyzed using the theory developed by Iordanski, Lyanda-Geller, and Pikus (ILP) for 2DES \cite{ILP}. A spin orbit length, $l_{so}$, of $\approx 0.76 \mu$m and coherence length, $l_{\phi}$, of $\approx 3 \mu$m are obtained. The fit obtained using the ILP model is plotted as a solid line in Fig.~2a. While the actual charge distribution of electrons in the quantum well is unknown during cool down,
band diagram calculations indicate that the charge distribution is not highly-asymmetric. This suggests that the Rashba spin-orbit coupling due to the electric field may be weak and spin-orbit terms due to biaxial strain (due to growth) and Dresselhaus contributions may be dominant \cite{Crooker_Strain_PRL}.  The fit suggests that in 2D, the spin splitting contour is isotropic in k-space and only one of the Rashba and Dresselhaus linear terms is dominant \cite{Winkler04}.


At $V_{sg} < -1.8$ V, when the wire is subjected to magnetic field, it exhibits negative magneto-resistance which ends with a pronounced kink at a characteristic field $B_{K}$, marking the field at which the classical cyclotron diameter equals the channel width \cite{Shabani08,vanHouten88}. Using $B_{K}$ one can then estimate the width of the channel: $w=2\hbar k_{F}/eB_{K}$, where $k_{F}$ is the Fermi wavevector. Figure 3d shows an example of magneto-resistance data measured at V$_{sg}$= -2.1 V. From the high magnetic field regime where SdH oscillations are dominant we deduce the carrier density and hence $k_{F}$. An effective width of $w \sim 190$nm is then determined from the kink position in Fig 2d. We repeat these measurements for several $V_{sg}$ and obtain the corresponding wire widths. Similar to 2D, we also study spin orbit parameters in our wire using the weak antilocalization signal in magnetotransport. An example of the weak antilocalization signal is shown in Fig.~2c.

\section*{Analysis and discussion}

In order to compare the spin orbit strength in our gate-defined wire with more stongly confined etched wires, we have fabricated single etched wires on similar InAs heterostructures. The wires were defined by electron beam lithography and dry etching. Ohmic contacts were made and the magneto transport measured as shown in Fig.~3a. Similar to the gate-defined wires, we determined the carrier density of the structure from the high magnetic field regime and using the kink position in magneto-resistance we obtain the width of the wire to be $w \sim 350$ nm. In Fig.~3b we observe geometrical resonances in magneto-resistance and a strong dip at zero magnetic field.

The spin orbit parameters are extracted using our quasi-1D model for conductivity at nonzero magnetic field \cite{Kettemann07,Kim13}:
\begin{align}
\delta \sigma_{1D}(B)=\frac{e^{2}}{h} &\bigg[\left(\frac{1}{l^2_\phi}+\frac{1}{l^2_B}\right)^{-\frac{1}{2}}
-\left(\frac{1}{l^2_\phi}+\frac{1}{l^2_B}+\frac{2r}{l^2_{so}}\right)^{-\frac{1}{2}} \nonumber \\
&-2\left(\frac{1}{l^2_\phi}+\frac{1}{l^2_B}+\frac{r}{l^2_{so}}\right)^{-\frac{1}{2}}\bigg],
\end{align}
where $r$ is a width-dependent coefficient that characterizes the effective strength of the spin-orbit coupling, $r = (w/l_{so})^{2}/12$ and $l_{B}$ is magnetic length. We also include the flux cancellation effect \cite{Beenakker88}. The fits obtained using Eq.~1 are shown in Fig.~2c and Fig.~3c. We have summarized the extracted spin-orbit parameters, $l_{so}$, $l_{\phi}$ in Fig.~4(a,b). The soft boundaries created by the electric field in our gate-defined wires lead to large values of $l_{\phi} \sim 2\mu$m. This is consistent with the fact that $l_{\phi}$ has a weak dependance on the wire width. In our etched wire, where the edges are exposed and edge roughness is introduced due to nanofabrication, $l_{\phi}$ drops to 60 \% of the gate-defined wires. However, the spin-orbit length is smaller due to stronger confinement.

Strong spin orbit interaction in our wires is reflected in small values of $l_{so}$ and is temperature indepedent. On the other hand, $l_{\phi}$ is a temperature-dependent quantity, affected by electron-electron interactions and boundary scattering in 1D.  For a fixed temperature, $l_{\phi}$ is related to the diffusion constant, $D$,  and dephasing time, $\tau_{\phi}$, through $l_{\phi} = \sqrt{D \tau_{\phi}}$. The diffusion constant depends on the amount of disorder present in the sample. In the quasi-1D limit, it is related to mean free path, $l_{e}$,  according to
$D = \frac{1}{2} v_{f} l_{e}$. Hence a large value of $l_{\phi}$ indicate less disorder and cleaner transport, for a fixed $\tau_\phi$.  

In the context of topological phases, spin orbit interaction sets the upper bound for the energy gap of the topological phase in a clean system. In a system where disorder is present, this gap is reduced and vanishes in highly disordered wires. In self-assembled nanowires very small values of spin orbit length ($\sim 100$nm) have been achieved due to the strong confinement. For typical InAs nanowires where surface is exposed, the mean free path has been also limited to $\sim 100$ nm mainly due to surface scattering \cite{Javey13}. Core-shell growth of InAs wires with surface coverage of a protective layer could in principle reduce the surface scattering centers and are currently under investigation \cite{LauhonCoreShellWire02, Marcus_Al_InAsWire}. Ideally large value of $l_{\phi}$ and small value of $l_{so}$ is needed for these systems and hence it seems natural to define the ratio $l_{\phi}/l_{so}$ as a figure of merit. Figure 4c shows the ratio $l_{\phi}/l_{so}$ as a function wire width.   We have also plotted data for InAs nanowires from Ref. \cite{HansenPRB05} taken at T = 0.2 K for comparison. Although the data is taken at higher temperature compared to our data, $l_{\phi} \sim 250$ nm shows a saturation for temperatures below 0.4 K.  From the data in Fig.~4c it is clear that less exposure of surfaces for one-dimensional channels is the key to maintain the quality and limit the surface scattering. 

In the work reported here, we have demonstrated that gate-defined InAs nanowires are a viable alternative to self-assembled vapor-liquid-solid InAs nanowire systems. Utilizing gate-controlled confinement not only preserves the quality of the wires but also offers greater tunability. Most importantly, the electronic properties are characterized in 2D before and after the formation of the gated-defined wires. This allows us to quantitatively compare the effect of disorder in gate- and etched-defined wires. The high quality of the wires and strong spin-orbit interaction characterized here show great promise for fabricating more complicated structures. Nanofabrication of gate-defined structures on these materials offers endless possibilities for exploring new directions. Our findings are expected to spark interest in large-scale device applications in spintronics as well as the search for Majorana states.

\textbf{Our work was supported by the Microsoft Research. A portion of this work was performed at the National High Magnetic Field Laboratory, which is supported by NSF Cooperative Agreement No. DMR-1157490, the State of Florida, and the U.S. DOE.}


\begin{thebibliography}{10}


\bibitem{Wilczek'09}
Wilczek, F.
\newblock Majorana returns.
\newblock {\em Nature Phys.}, {\bf 5}, 614--618, (2009).


\bibitem{Stern'10}
Stern, A.
\newblock Majorana returns.
\newblock {\em Nature}, {\bf 464}, 187-193, (2010).


\bibitem{Franz'10}
Franz, M.
\newblock Race for majorana fermions.
\newblock {\em Physics}, {\bf 3}, 24, (2010).


\bibitem{Nayak'10}
Nayak, C.
\newblock Majorana returns.
\newblock {\em Nature}, {\bf 464}, 693--694, (2010).


\bibitem{Kane_review}
Hasan, M.~Z. and Kane, C.~L.
\newblock Colloquium: Topological insulators.
\newblock {\em Rev. Mod. Phys.}, {\bf 82}, 3045-3067, (2010).

\bibitem{Qi_review}
Qi, X.-L. and Zhang, S.-C. 
\newblock Topological insulators and superconductors.
\newblock {\em Rev. Mod. Phys.}, {\bf 83}, 1057-1110, (2011).

\bibitem{Alicea12}
Alicea, J.
\newblock New directions in the pursuit of majorana fermions in solid state
  systems.
\newblock {\em Reports on Progress in Physics}, {\bf 75}, 076501, (2012).

\bibitem{Beenakker_review}
Beenakker, C.~W.~J.
\newblock Search for majorana fermions in superconductors.
\newblock {\em Annual Review of Condensed Matter Physics}, {\bf 4}, 113-136, (2013).


\bibitem{Sau10}
Sau, J.~D., Lutchyn, R.~M., Tewari S., and Das Sarma, S.
\newblock Generic new platform for topological quantum computation using
  semiconductor heterostructures.
\newblock {\em Phys.\ Rev.\ Lett.}, {\bf 104}, 040502, (2010).


\bibitem{Alicea10}
Alicea, J.
\newblock Majorana fermions in a tunable semiconductor device.
\newblock {\em Phys.\ Rev.\ B}, {\bf 81}, 125318, (2010).

\bibitem{Lutchyn10}
Lutchyn, R.~M., Sau J.~D., and Das~Sarma, S.
\newblock Majorana fermions and a topological phase transition in
  semiconductor-superconductor heterostructures.
\newblock {\em Phys. Rev. Lett.}, {\bf 105}, 077001, (2010).


\bibitem{Oreg10}
Oreg, Y., Refael, G. and von Oppen, F.
\newblock Helical liquids and majorana bound states in quantum wires.
\newblock {\em Phys. Rev. Lett.}, {\bf 105}, 177002,  (2010).



\bibitem{NayakRevModPhys}
Nayak, C., Simon, S.~H., Stern, A., Freedman, M. and 
  Das~Sarma, S..
\newblock Non-abelian anyons and topological quantum computation.
\newblock {\em Rev. Mod. Phys.}, {\bf 80}, 1083, (2008).


\bibitem{Mourik2012}
Mourik, V., et al.
\newblock {Signatures of Majorana Fermions in Hybrid
  Superconductor-Semiconductor Nanowire Devices}.
\newblock {\em Science}, {\bf 336}, 1003, (2012).



\bibitem{Rokhinson2012}
Rokhinson, L.~P., Liu, X. and Furdyna, J.~K.
\newblock Observation of the fractional ac josephson effect: the signature of
  majorana particles.
\newblock {\em Nature Phys.}, {\bf 8}, 795, (2012).

\bibitem{Das2012}
Das, A., et al.
\newblock {Zero-bias peaks and splitting in an Al-InAs nanowire topological
  superconductor as a signature of Majorana fermions}.
\newblock {\em Nature Phys.}, {\bf 8}, 887, (2012).

\bibitem{Deng2012}
Deng, M.~T.,  et al.
\newblock {Observation of Majorana Fermions in a Nb-InSb Nanowire-Nb Hybrid
  Quantum Device}.
\newblock {\em Nano Lett.}, {\bf 12}, 6414, (2012).


\bibitem{Fink2012}
Finck, A.~D.~K.,  et al.
\newblock {Anomalous modulation of a zero bias peak in a hybrid
  nanowire-superconductor device}.
\newblock {\em Phys. Rev. Lett.}, {\bf 110}, 126406, (2013).

\bibitem{Churchill2013}
Churchill, H.~O.~H., et al.
\newblock Superconductor-nanowire devices from tunneling to the multichannel
  regime: Zero-bias oscillations and magnetoconductance crossover.
\newblock {\em Phys. Rev. B}, {\bf 87}, 241401, (2013).


\bibitem{AliceaBraiding}
 Alicea, J., et al.
\newblock Non-abelian statistics and topological quantum information processing
  in 1d wire networks.
\newblock {\em Nature Phys.}, {\bf 7} 412, (2011).



\bibitem{HansenPRB05}
Hansen, A.~E., et al. 
\newblock Spin relaxation in inas nanowires studied by tunable weak
  antilocalization.
\newblock {\em Phys. Rev. B}, {\bf 71}, 205328, (2005).



\bibitem{SchaperPRB06}
Sch\"apers, T., et al.
\newblock Suppression of weak antilocalization in
  \uppercase{G}a$_{x}$\uppercase{I}n$_{1-x}$\uppercase{A}s/\uppercase{I}n\uppercase{P}
  narrow quantum wires.
\newblock {\em Phys. Rev. B}, {\bf 74}, 081301, (2006).


\bibitem{Kunihashi10}
Kunihashi, Y., Kohada, M., and Nitta, J.
\newblock Experimental demonstration of resonant spin-orbit interaction effect.
\newblock {\em Physics Procedia}, {\bf 3}, 1255, (2010).


\bibitem{RoulleauPRB10}
Roulleau, P., et al.
\newblock Suppression of weak antilocalization in \uppercase{i}n\uppercase{a}s nanowires.
\newblock {\em Phys. Rev. B}, {\bf 81}, 155449, (2010).



\bibitem{KallaherPRB13}
Kallaher, R.~L.,Heremans,  J.~J., van~Roy, W., and Borghs, G..
\newblock Spin and phase coherence lengths in inas wires with diffusive
  boundary scattering.
\newblock {\em Phys. Rev. B}, {\bf 88}, 205407, (2013).



\bibitem{Sau_review}
Sau, J.~D., et al.
\newblock Non-abelian quantum order in spin-orbit-coupled semiconductors:
  Search for topological majorana particles in solid-state systems.
\newblock {\em Phys. Rev. B}, {\bf 82}, 214509, (2010).


\bibitem{Motrunich01_Disorder_in_topological_1D_SC}
Motrunich, O., Damle, K., and Huse, D.~A.
\newblock Griffiths effects and quantum critical points in dirty
  superconductors without spin-rotation invariance: One-dimensional examples.
\newblock {\em Phys. Rev. B}, {\bf 63}, 224204, (2001).



\bibitem{Gruzberg05_Localization_in_disordered_SC_wires_with_broken_SU2_symmetry}
Gruzberg, I.~A., Read, N., and Vishveshwara, S.
\newblock Localization in disordered superconducting wires with broken
  spin-rotation symmetry.
\newblock {\em Phys. Rev. B}, {\bf 71}, 245124, (2005).

\bibitem{Brouwer11_Topological_SC_in_disorder_wires}
Brouwer, P.~W., Duckheim, M., Romito, A., and von Oppen, F.
\newblock Topological superconducting phases in disordered quantum wires with
  strong spin-orbit coupling.
\newblock {\em Phys. Rev. B}, {\bf 84}, 144526, (2011).


\bibitem{LobosPRL12}
 Lobos, A.~M., Lutchyn, R.~M. and Das~Sarma, S.
\newblock Interplay of disorder and interaction in majorana quantum wires.
\newblock {\em Phys. Rev. Lett.}, {\bf 109}, 146403, (2012).


\bibitem{Sau12}
Sau, J.~D., Tewari, S., and Das~Sarma, S.
\newblock Experimental and materials considerations for the topological
  superconducting state in electron- and hole-doped semiconductors: Searching
  for non-abelian majorana modes in 1d nanowires and 2d heterostructures.
\newblock {\em Phys. Rev. B}, {\bf 85}, 064512, (2012).



\bibitem{Petta05}
Petta, J.~R., et al. 
\newblock Coherent manipulation of coupled electron spins in semiconductor
  quantum dots.
\newblock {\em Science}, {\bf 309}, 2180, (2005).


\bibitem{Heiblum}
Neder, I., et al.
\newblock Interference between two indistinguishable electrons from independent
  sources.
\newblock {\em Nature}, {\bf 448}, 333, (2007).


\bibitem{Willett09a}
Willett, R.~L., Pfeiffer,  L.~N. and West, K.~W.
\newblock Measurement of filling factor 5/2 quasiparticle interference with
  observation of charge e/4 and e/2 period oscillations.
\newblock {\em PNAS}, {\bf 106}, 8853, (2009).

\bibitem{Kroemer04}
Kroemer, H.
\newblock The 6.1\AA family (\uppercase{i}n\uppercase{a}s, \uppercase{g}a\uppercase{s}b, \uppercase{a}l\uppercase{s}b) and its heterostructures: a
  selective review.
\newblock {\em Physica E: Low-dimensional Systems and Nanostructures}, {\bf 20}, 196, (2004).
\newblock Proceedings of the 11th International Conference on Narrow Gap
  Semiconductors.


\bibitem{Capotondi04}
Capotondi, F., et al.
\newblock {\em J. Vac. Sci Tech B}, {\bf 22}, 702, (2004).


\bibitem{Richter00}
Richter, A., et al.
\newblock Transport properties of modulation-doped inas-inserted-channel
  inalas/ingaas structures grown on gaas substrates.
\newblock {\em Applied Physics Letters}, {\bf 77}, 3227, (2000).


\bibitem{Wallart05}
Wallart, X., Lastennet, J., Vignaud, D., and Mollot, F..
\newblock Performances and limitations of  \uppercase{I}n\uppercase{A}s/\uppercase{I}n\uppercase{A}l\uppercase{A}s metamorphic
  heterostructures on inp for high mobility devices.
\newblock {\em Applied Physics Letters}, {\bf 87}, 043504, (2005).


\bibitem{DasSarmaHwangPRB13}
Das~Sarma, S. and Hwang, E.~H.
\newblock Universal density scaling of disorder-limited low-temperature
  conductivity in high-mobility two-dimensional systems.
\newblock {\em Phys. Rev. B}, {\bf 88}, 035439, (2013).


\bibitem{vanWees88}
van Wees, B.~J., et al.
\newblock Quantized conductance of point contacts in a two-dimensional electron
  gas.
\newblock {\em Phys. Rev. Lett.}, {\bf 60}, 848, (1988).


\bibitem{Wharam88}
Wharam, D.~A., et al.
\newblock {\em J. Phys. C}, {\bf 21}, L209, (1988).



\bibitem{SiGe2}
Wieser, U., Kunze, U., Ismail, K. and Chu, J.~O. 
\newblock Quantum-ballistic transport in an etch-defined \uppercase{s}i/\uppercase{s}i\uppercase{g}e quantum point contact.
\newblock {\em Applied Physics Letters}, {\bf 81}, 1726, (2002).



\bibitem{GaN}
Chou, H.~T., et al.
\newblock High-quality quantum point contacts in \uppercase{g}a\uppercase{n}/\uppercase{a}l\uppercase{g}a\uppercase{n} heterostructures.
\newblock {\em Applied Physics Letters}, {\bf 86}, 073108, (2005).




\bibitem{AlAs}
Gunawan, O., Habib, B., De~Poortere, E.~P., and Shayegan, M.
\newblock Quantized conductance in an alas two-dimensional electron system
  quantum point contact.
\newblock {\em Phys. Rev. B}, {\bf 74}, 155436, (2006).


\bibitem{Winkler03}
Winkler, R.
\newblock {\em Spin-Orbit Coupling Effects in Two-Dimensional Electron and Hole
  Systems}.
\newblock Springer Berlin, 2003.


\bibitem{Koester96}
Koester, S.~J., et al. 
\newblock Length dependence of quantized conductance in ballistic constrictions
  fabricated on \uppercase{i}n\uppercase{a}s/\uppercase{a}l\uppercase{s}b quantum wells.
\newblock {\em Phys. Rev. B}, {\bf 53}, 13063, (1996).



\bibitem{Cronenwett02}
 Cronenwett, S.~M, et al.
\newblock Low-temperature fate of the 0.7 structure in a point contact: A
  kondo-like correlated state in an open system.
\newblock {\em Phys. Rev. Lett.}, {\bf 88}, 226805, (2002).


\bibitem{dePicciotto05}
de~Picciotto, R., Pfeiffer, L.~N.,  Baldwin, K.~W. and West, K.~W. 
\newblock Temperature-dependent 0.7 structure in the conductance of
  cleaved-edge-overgrowth one-dimensional wires.
\newblock {\em Phys. Rev. B}, {\bf 72}, 033319, (2005).


\bibitem{ILP}
Iordanskii, S.~V. , Lyanda-Geller, Y.~B. and Pikus, G.~E. 
\newblock {\em JETP Lett.}, {\bf 60}, 206, (1994).



\bibitem{Crooker_Strain_PRL}
Crooker, S.~A. and Smith, D.~L.
\newblock Imaging spin flows in semiconductors subject to electric, magnetic,
  and strain fields.
\newblock {\em Phys. Rev. Lett.}, {\bf 94}, 236601, (2005).



\bibitem{Winkler04}
Winkler, R.
\newblock Spin orientation and spin precession in inversion-asymmetric
  quasi-two-dimensional electron systems.
\newblock {\em Phys. Rev. B}, {\bf 69}, 045317, (2004).



\bibitem{Shabani08}
Shabani, J., Petta, J.~R., and Shayegan, M.
\newblock High-quality quantum point contact in two-dimensional \uppercase{g}a\uppercase{a}s (311)\uppercase{a}
  hole system.
\newblock {\em Applied Physics Letters}, {\bf 93}, 212101, (2008).



\bibitem{vanHouten88}
van Houten, H., et al.
\newblock Four-terminal magnetoresistance of a two-dimensional electron-gas
  constriction in the ballistic regime.
\newblock {\em Phys. Rev. B}, {\bf 37}, 8534, (1988).


\bibitem{Kettemann07}
Kettemann, S.
\newblock Dimensional control of antilocalization and spin relaxation in
  quantum wires.
\newblock {\em Phys. Rev. Lett.}, {\bf 98}, 176808, (2007).



\bibitem{Kim13}
Kim, Y., Lutchyn,  R.~M., and Nayak, C.
\newblock Origin and transport signatures of spin-orbit interactions in one-
  and two-dimensional \uppercase{S}r\uppercase{T}i\uppercase{O}$_{3}$-based
  heterostructures.
\newblock {\em Phys. Rev. B}, {\bf 87}, 245121, (2013).


\bibitem{Beenakker88}
Beenakker, C.~W.~J.  and van Houten,  H.
\newblock Flux-cancellation effect on narrow-channel magnetoresistance
  fluctuations.
\newblock {\em Phys. Rev. B}, {\bf 37}, 6544, (1988).


\bibitem{Javey13}
Chuang, S., et al.
\newblock Ballistic \uppercase{i}n\uppercase{a}s nanowire transistors.
\newblock {\em Nano Letters}, {\bf 13}, 555 (2013).



\bibitem{LauhonCoreShellWire02}
Lauhon, L.~J. , Gudiksen, M.~S. , Wang, D., and Lieber, C.~M. .
\newblock Epitaxial core-shell and core-multishell nanowire heterostructures.
\newblock {\em Nature}, {\bf }, {\bf 420}, 57, (2002).


\bibitem{Marcus_Al_InAsWire}
Ziino, N.~L.~B., et al.
\newblock Epitaxial aluminum contacts to \uppercase{i}n\uppercase{a}s nanowires.
\newblock {\em ar\uppercase{x}iv: 1309.4569}, 2014.

\end{thebibliography}

\begin{figure*}[htp]
\centering
\includegraphics[scale=0.75]{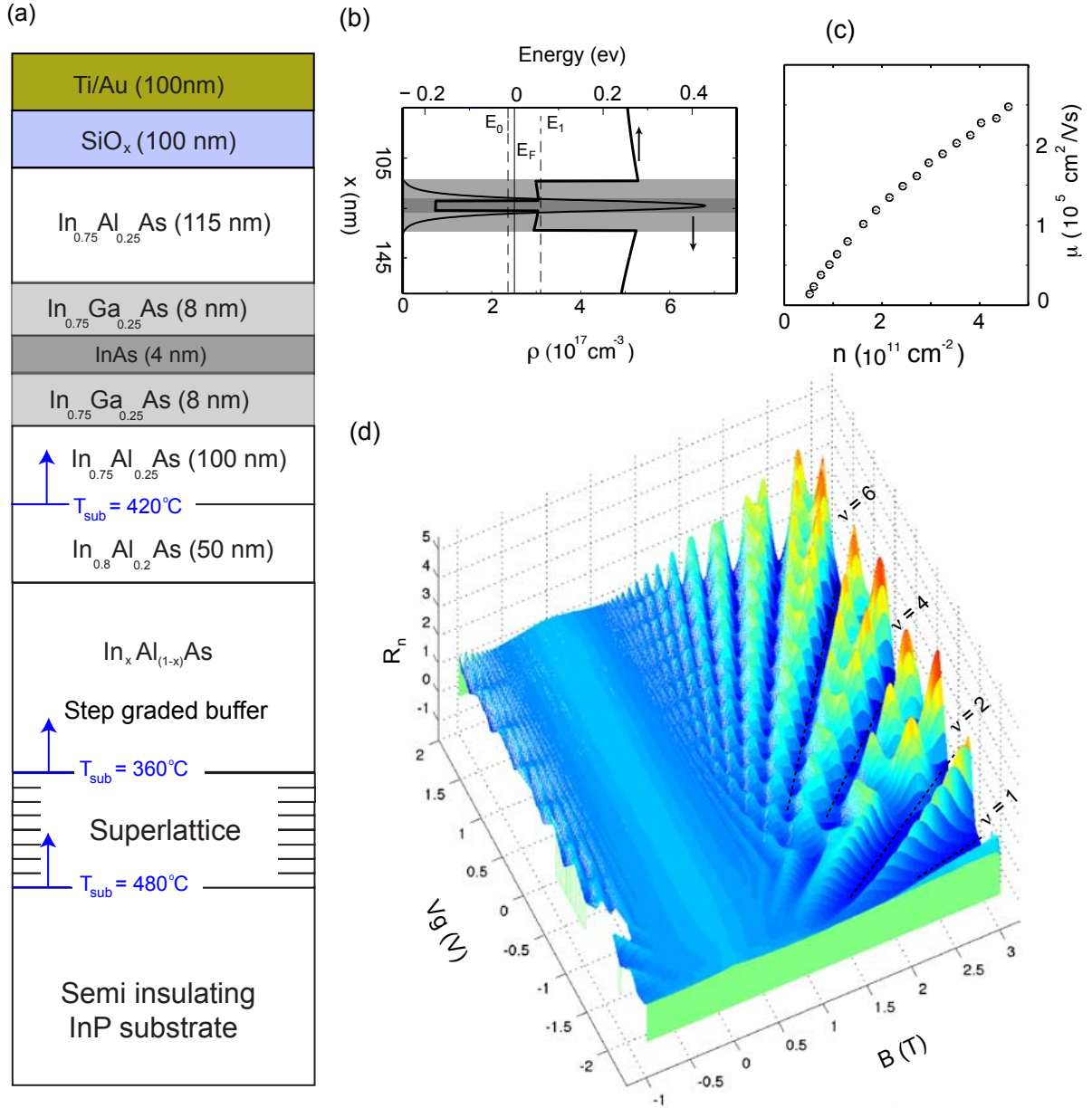}
    \caption{(Color online)  (a) Epitaxial structure of MBE grown InAs quantum well system capped with SiO$_{x}$ and top gate with corresponding band diagram and charge distribution of the active region shown in (b). (c) Mobility measured in a gated-Hall bar as a function of density. (d) 3D plot of normalized magneto-resistance traces ($R_{n}$ = R(B)/R(B = 0)) as a function of gate voltage.}
\end{figure*}

\begin{figure*}[htp]
\centering
\includegraphics[scale=0.75]{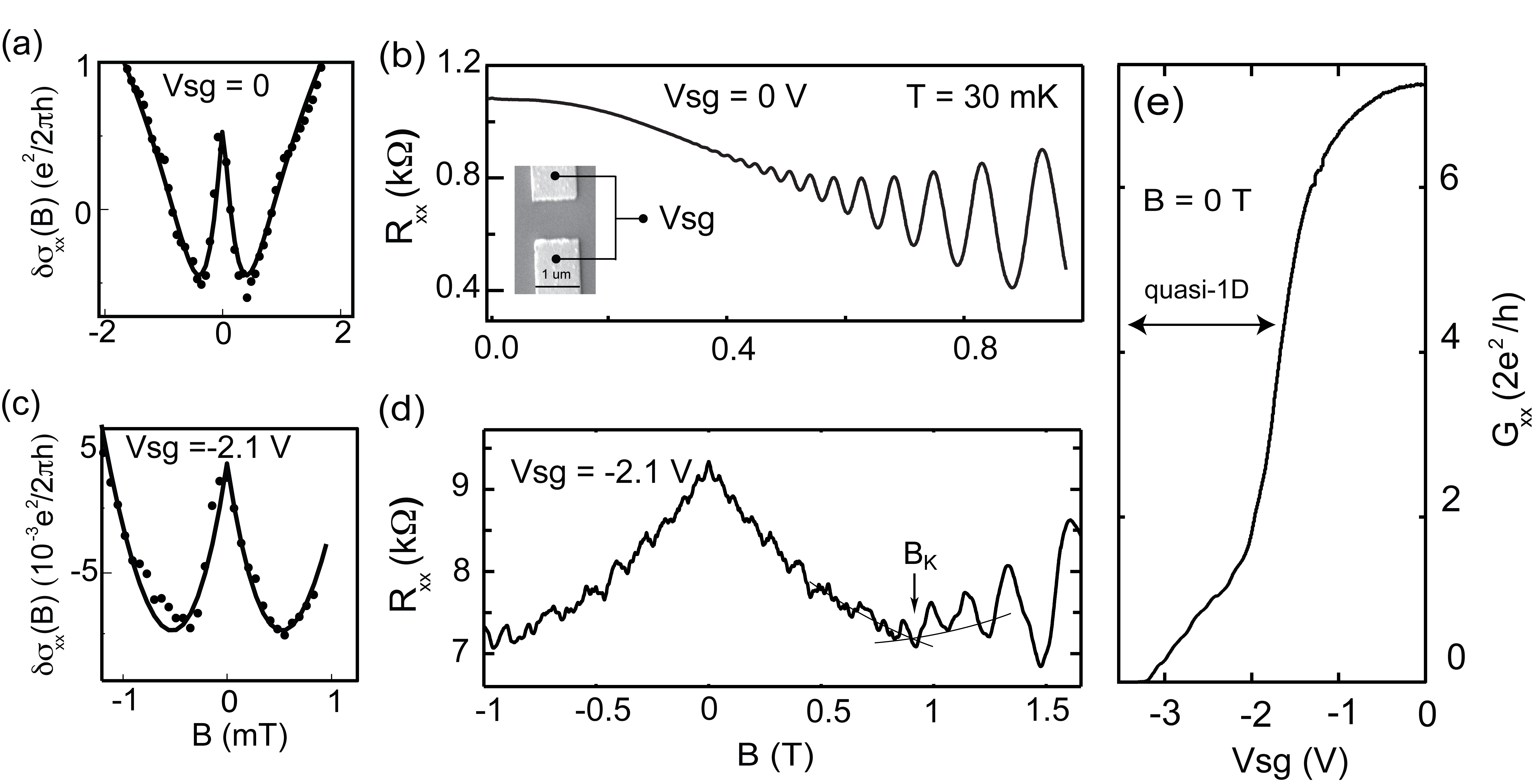}
    \caption{(Color online)  (a) Weak antilocalization signal at $V_{sg}$ = 0 V in magneto-conductance near zero magnetic field. The fit obtained using the ILP model for 2DEG is shown in solid curve. (b) Shubnikov-de Haas oscillations at $V_{sg}$ = 0 V. Inset: SEM image of the split gate device. (c) same as (a) at $V_{sg}$ = -2.1 V. The fit obtained using Eq.~1 is shown in solid curve. (d) magneto-resistance through the constriction at $V_{sg}$ = -2.1 V. Shubnikov de Haas oscillations appear at $B_{K}$. (e) Plot of conductance vs side gate bias with B = 0 T. The sharp drop in conductance indicates depletion of electrons under the side gates $V_{sg} \sim$ -1.8 V. }
\end{figure*}

\begin{figure*}[htp]
\centering
\includegraphics[scale=0.75]{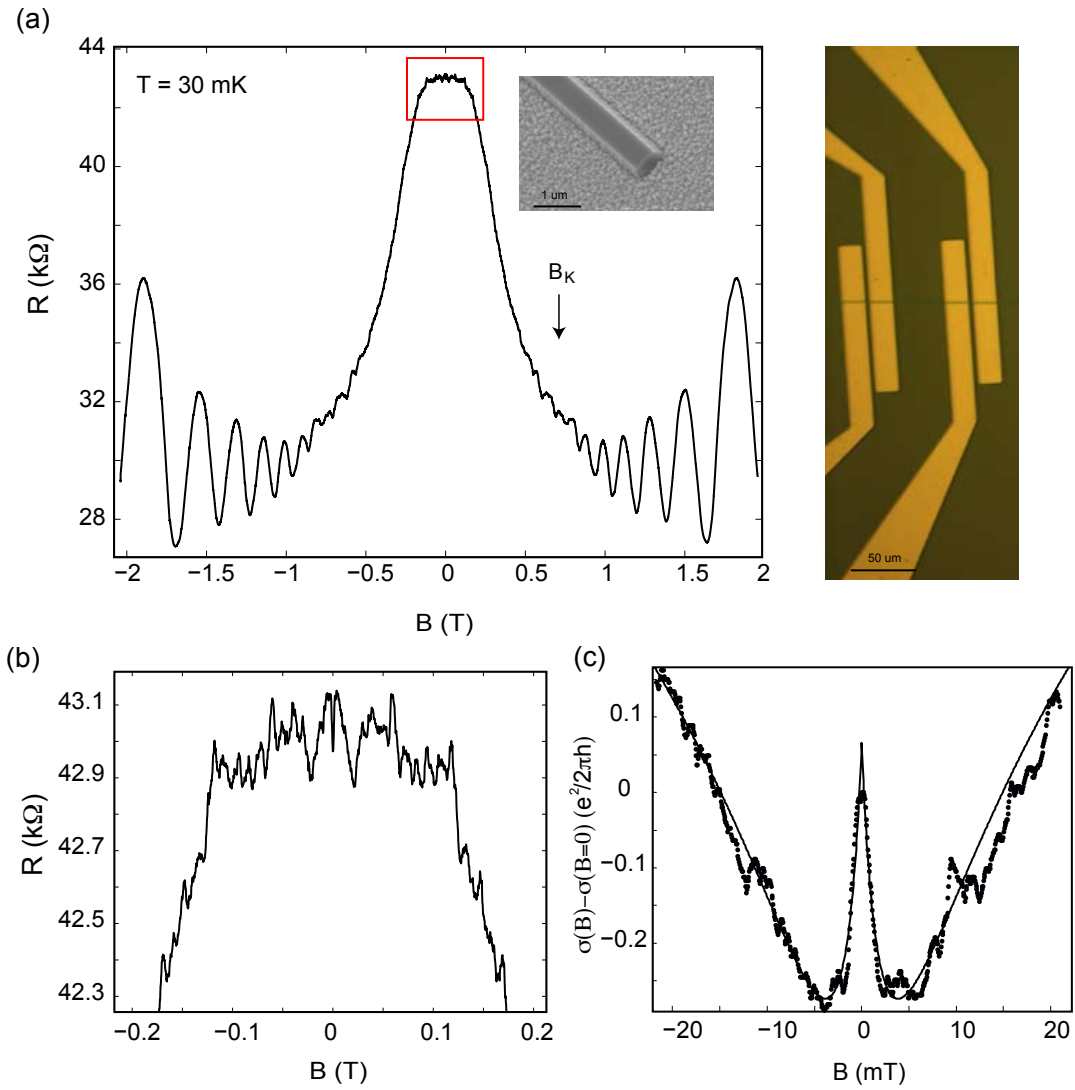}
    \caption{(Color online)  (a) Magneto-resistance data measured on a single etched wire. Inset shows an SEM image of the wire. The wire is 200 $\mu$m long. (b) Low magnetic field region (indicated by red box in (a)) shows signature of geometrical resonances and a strong dip at zero magentic field. (c) The weak antilocalization signal in magneto-conductance is shown together with the fit (solid curve). }
\end{figure*}

\begin{figure*}[htp]
\centering
\includegraphics[scale=0.825]{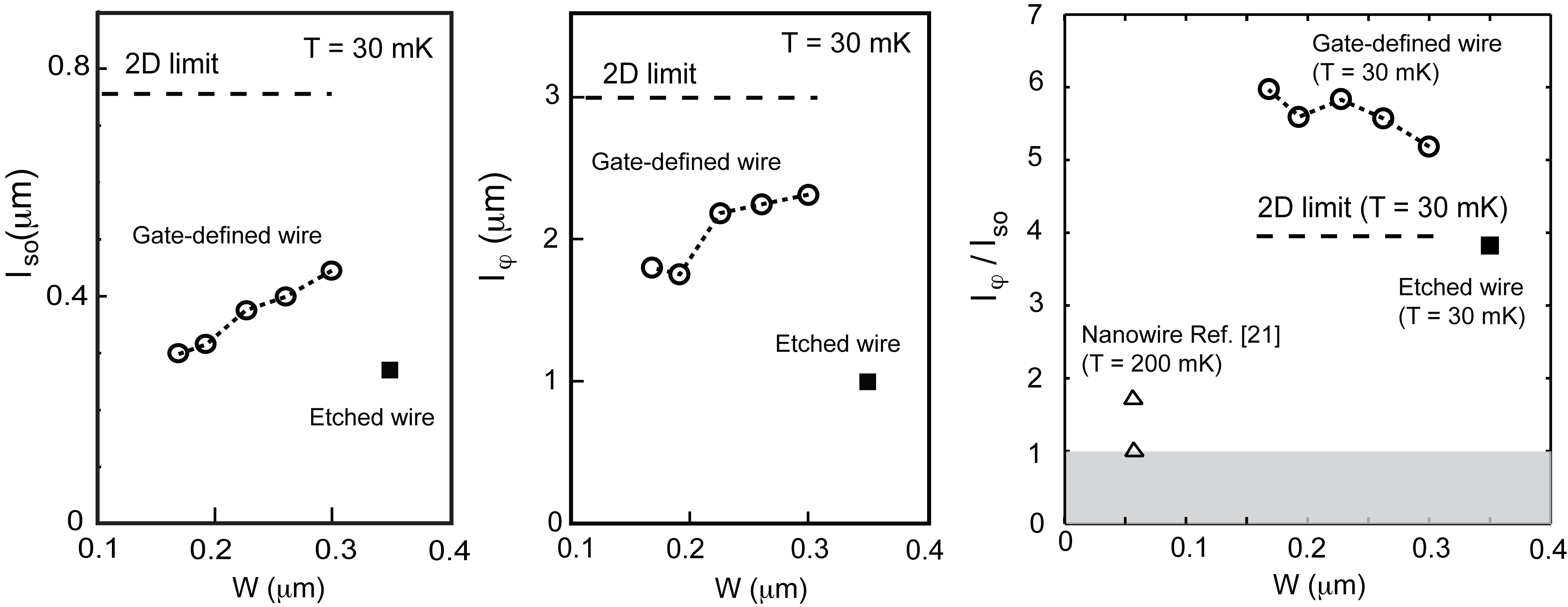}
    \caption{(Color online)  (a) Spin orbit lengths and (b) coherent lengths extracted from weak antilocalization measurements for gate-defined and etched wires plotted as a function of wire width. The dashed lines shows the values extracted at $V_{sg}$ = 0 V (2D limit). (c) The ratio of $l_{\phi}/l_{so}$ is plotted from (a) and (b). For comparison we have plotted data taken in InAs self-assembled nanowires from Ref. \cite{HansenPRB05}. }
\end{figure*}

\end{document}